\documentclass[twocolumn,showkeys,prc,nofootinbib]{revtex4-1}

\usepackage{amsmath}
\usepackage{amssymb}
\usepackage{graphicx}
\usepackage{epsfig}
\usepackage{natbib}

\begin{document}

\title{Thermal evolution of neutron stars with global and local neutrality}

\author{S. M. de Carvalho$^{1,2}$, R.~Negreiros$^{2}$, Jorge A. Rueda$^{1,3,4}$, Remo Ruffini$^{1,3,4}$}
\affiliation{$^1$ICRANet-Rio, Centro Brasileiro de Pesquisas F\'isicas, Rua Dr. Xavier Sigaud 150, Rio de Janeiro, RJ, 22290-180, Brazil }
\affiliation{$^2$Instituto de F\'isica, Universidade Federal Fluminense, UFF, Niter\'oi, 24210-346, RJ, Brazil}
\affiliation{$^3$Dipartimento di Fisica and ICRA, Sapienza Universit\`a di Roma, P.le Aldo Moro 5, I--00185 Rome, Italy}
\affiliation{$^4$ICRANet, P.zza della Repubblica 10, I--65122 Pescara, Italy}

\date{\today}

\begin{abstract}

Globally neutral neutron stars, obtained from the solution of the called 
Einstein-Maxwell-Thomas-Fermi equations that account for all the fundamental 
interactions, have been recently introduced. These configurations have a more 
general character than the ones obtained with the traditional 
Tolman-Oppenheimer-Volkoff, which impose the condition of local charge 
neutrality. The resulting configurations have a less massive and thinner crust, 
leading to a new mass-radius relation. Signatures of this new structure of the 
neutron star on the thermal evolution might be a potential test for this theory. 
We compute the cooling curves by integrating numerically the energy balance and 
transport equations in general relativity, for globally neutral neutron stars 
with crusts of different masses and sizes, according to this theory for 
different core-crust transition interfaces. We compare and contrast our study 
with known results for local charge neutrality. We found a new behavior for the 
relaxation time, depending upon 
the density at the base of the crust, $\rho_{\rm crust}$. In particular, we find 
that the traditional increase of the relaxation time with the crust thickness 
holds only for configurations whose density of the base of the crust is greater 
than $\approx 5\times 10^{13}$~g~cm$^{-3}$. The reason for this is that neutron 
star crusts with very thin or absent inner crust have some neutrino emission 
process blocked which keep the crust hotter for longer times. Therefore, 
accurate observations of the thermal relaxation phase of neutron stars might 
give crucial information on the core-crust transition which may aid us in 
probing the inner composition/structure of these objects.

\end{abstract}


\maketitle
%
\section{Introduction}\label{sec:1}

In recent works \citep{rotondo11d,rueda11,belvedere12,2014NuPhA.921...33B} a new 
approach has been developed in which a neutron star is considered to have global 
rather than local charge neutrality. It was shown that the new equilibrium 
equations, the Einstein-Maxwell-Thomas-Fermi (EMTF) equations, introduce 
self-consistently the presence of the electromagnetic interactions in addition 
to the strong, weak, and gravitational interactions, all within the framework of 
general relativity. The weak interactions are introduced by the 
$\beta$-stability, and the strong interactions are modeled via the 
$\sigma$-$\omega$-$\rho$ nuclear model. In this work we adopt the NL3 
parameterization of this nuclear model \citep[see][and references therein for 
more details]{belvedere12}. The supranuclear core is composed by a degenerate 
gas of neutrons, protons, and electrons. The crust, in its outer region ($\rho 
\leq \rho_{\rm drip}\approx 4.3\times 10^{11}$ g~cm$^{-3}$), is composed of ions 
and electrons, while in its inner 
region ($\rho_{\rm drip}<\rho<\rho_{\rm nuc}$, where $\rho_{\rm nuc}\approx 
2.7\times 10^{14}$~g~cm$^{-3}$ is the nuclear saturation density), there is an 
additional component of free neutrons dripped out from nuclei.

\begin{figure*}[!hbtp]
\centering\includegraphics[width=\hsize,clip]{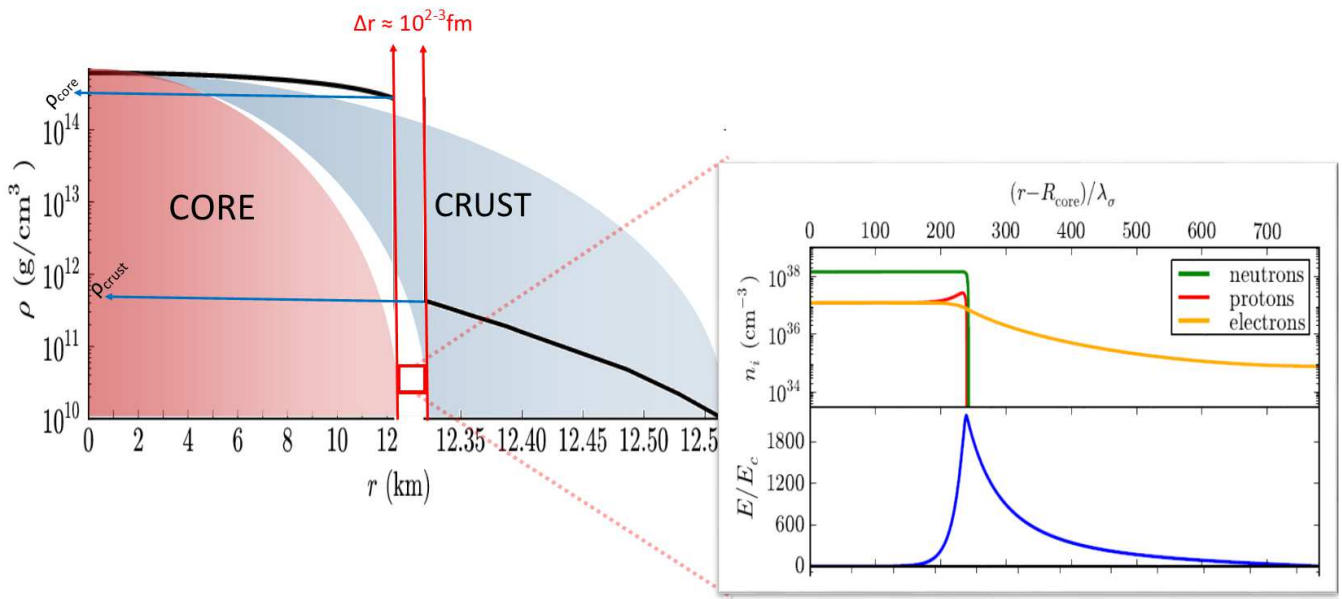}
\caption{In the top and center panels we show the neutron, proton, electron 
densities and the electric field in units of the critical electric field $E_c$ 
in the core-crust transition layer, whereas in the bottom panel we show a 
specific example of a density profile inside a neutron star. In this plot we 
have used for the globally neutral case a density at the edge of the crust equal 
to the neutron drip density, $\rho_{\rm drip}\approx 4.3\times 
10^{11}$~g~cm$^{-3}$, and $\lambda_{\sigma}=\hbar/(m_{\sigma}c)\sim 0.4$~fm 
denotes the $\sigma$-meson Compton wavelength.}\label{fig:Model}
\end{figure*}

The solution of the EMTF equations leads to a new structure of neutron stars, 
significantly different from the traditional configurations obtained through the 
TOV equations (see Fig.~\ref{fig:Model}): the core is positively charged due to 
gravito-polarization, giving rise to a Coulomb potential energy, $e V\sim m_\pi 
c^2$. The core-crust transition takes place at the nuclear saturation density 
$\rho=\rho_{\rm nuc}$. The transition is signaled by the existence of a a thin 
($\Delta r\sim$few hundreds fm) electron layer, fully screening the core charge. 
In this transition layer the electric field becomes overcritical, $E\sim 
(m_\pi/m_e) E_c$ with $E_c =m_e^2 c^3/(e \hbar)$ the critical field for vacuum 
polarization, and the particle densities decrease until the base of the crust, 
which is reached when global charge neutrality is achieved. Consequently, the 
density at the base of the crust is characterized by $\rho_{\rm crust}\leq 
\rho_{\rm nuc}$. Furthermore 
due to the appearance of an electric field, the core-crust transition is no 
longer contiguous, since as discussed above there is now a gap with width 
$\Delta r$ filled by the screening electrons.

\begin{figure}[!hbtp]
\centering\includegraphics[width=\hsize,clip]{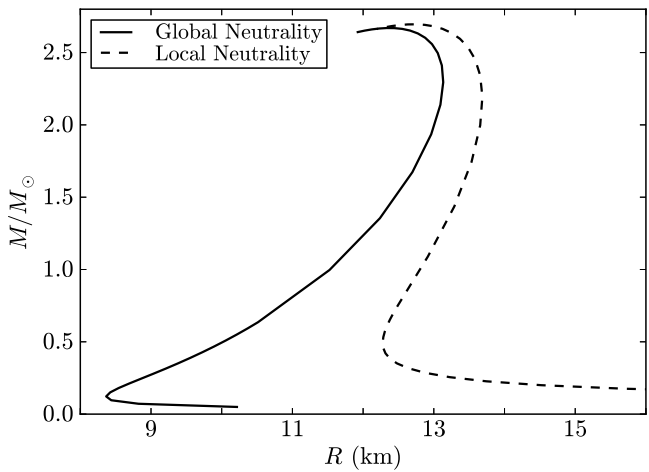}
\caption{Mass-radius relation obtained with the traditional locally neutral TOV 
treatment case and the global charge neutrality configurations, with $\rho_{\rm 
crust}=\rho_{\rm drip}$ \cite{belvedere12}. Configurations lying between the 
solid and dashed curves have $\rho_{\rm crust} >\rho_{\rm drip}$ and so they 
possess inner crust.}
\label{fig:fig1}
\end{figure}

Configurations with $\rho_{\rm crust}>\rho_{\rm drip}$ possess both inner and 
outer crust, whereas in the cases with $\rho_{\rm crust}\leq \rho_{\rm drip}$ 
the neutron star has only an outer crust. In the limit $\rho_{\rm 
crust}\to\rho_{\rm nuc}$, both $\Delta r$ and $E$ of the transition layer 
vanish, and the solution approaches the one given by local charge neutrality 
(see Figs.~3 and 5 in \cite{belvedere12}). Details on the boundary and 
equilibrium conditions leading to these configurations are presented below in 
the next section. All the above features lead to a new mass-radius relation of 
neutron stars as shown in Fig.~\ref{fig:fig1}; see \cite{belvedere12} for 
further details.

The aim of this work is to compute the thermal evolution of globally neutral 
neutron stars for different values of the density of the crust, $\rho_{\rm 
crust}$, covering configurations with and without inner crust, all the way to 
the limit $\rho_{\rm crust}\approx\rho_{\rm nuc}$, which corresponds to TOV-like 
solutions satisfying local charge neutrality. The article is organized as 
follows. We discuss in section \ref{sec:2} gravito-polarization effects in 
neutron stars and the boundary conditions of neutron stars satisfying global 
charge neutrality. In section \ref{sec:3} we recall the equations of thermal 
evolution in the spherically symmetric case and the cooling mechanisms taken 
into account in this work. We show in section \ref{sec:4} the cooling curves 
obtained from the numerical integration of the thermal evolution equations and 
compute the thermal relaxation time of the configurations. The effects of 
pairing effects in the crust in the cooling curves are shown in section 
\ref{sec:5} and we finally summarize our conclusions in section \ref{sec:6}. We 
use geometric units $G=c=1$ throughout the article.

\section{Core-crust transition and family of globally neutral 
solutions}\label{sec:2}

The specific value of $\rho_{\rm crust}$ establishes the properties of the 
core-crust transition surface as follows. The matching of the core to the crust 
is at $r=R_{\rm core} + \Delta r$. The equilibrium conditions are given by the 
constancy of the particle Klein potentials (or generalized Fermi energies) 
throughout the configuration
\begin{equation}\label{eq:Efermi}
E^F_i =e^{\nu(r)/2}\mu_i(r) + \mathcal{V}_i(r) = {\rm constant},
\end{equation}
where $\mu_i$ and $\mathcal{V}_i$ are the free-chemical potential and effective 
potentials of the $i$-specie. $\mathcal{V}_i$ includes the electromagnetic and 
nuclear potentials. $e^{\nu(r)}$ is the 0--0 component of the spherically 
symmetric metric 
\begin{equation}\label{eq:metric}
ds^2=e^{\nu(r)} dt^2 - \frac{dr^2}{1-2m(r)/r} -r^2 d\theta^2 - r^2 \sin^2\theta 
d\varphi^2,
\end{equation}
with $r$ the radial coordinate and $m(r)$ is the mass function.

The matter in the crust is thought to be solid-like composed of an 
electron-ionic lattice at densities $\rho\leq\rho_{\rm drip}$ and by an 
electron-neutron-ionic lattice at $\rho>\rho_{\rm drip}$. The description of the 
matter can be simplified by the introduction of Wigner-Seiz cells: each cell 
contains a central ion of total positive charge $+eZ$ ($Z$ denotes the number of 
protons), surrounded by compressed degenerate cloud of relativistic electrons 
which fully screened the ion positive charge at the cell's border (i.e.~number 
of electrons in the cell equal to $Z$). At densities higher than the 
neutron-drip value, there is present also the background of neutrons. The 
equation of state of matter in such a state of high density and pressure was 
first computed in a classic work by Baym, Bethe and Pethick \cite{baym71a}. The 
distribution of electrons around each nucleus is assumed to be uniform and the 
nuclear model was based on a phenomenological extension of the semi-empirical 
liquid-drop model which allows nuclear compressibility. It was also included the 
reduction of the nuclear surface tension in presence of the neutron background 
for densities higher than the neutron-drip value. The effect of inhomogeneities 
in the electron distribution around the nucleus was recently examined in 
\cite{rotondo11b} by extending the classic Thomas-Fermi model to relativistic 
regimes and introducing the nucleus finite size. The extension to finite 
temperatures of that treatment was recently presented in 
\cite{2014PhRvC..89a5801D}. The electron density, so computed, takes into 
account self-consistently both the electron-electron and electron-ion Coulomb 
interactions. The equilibrium configuration of the electrons is characterized by 
higher values close to the nucleus surface and decreases outwards until it fully 
screens the ion at the Wigner-Seitz cell's radius. 

The Wigner-Seitz cells, which are globally neutral, can be thought as the 
building blocks of the crust. The equilibrium distribution of such neutral cells 
in the crust is then expected to occur without gravito-polarization, and 
therefore no net electric field, Coulomb potential, and charge should appear. On 
the contrary, in the neutron star core, charge separation occurs since charged 
particles (proton and electrons) are `free' to flow, gravitationally 
segregating. Indeed, as shown in Eq.~(21) in \cite{rotondo11d}, the Coulomb 
potential depth at the center of the core is proportional to the proton-electron 
mass difference, i.e.~$eV(r=0)\propto (m_p-m_e)c^2$. Hence, a transition from a 
charged region (the core) to a neutral region (the crust), has to be present in 
the system. 

For the electron gas, the Eq.~(\ref{eq:Efermi}) imposes, in the transition 
layer, the following continuity condition
\begin{equation}\label{eq:Efermicrust}
E^F_e = e^{\nu_{\rm core}/2}\mu^{\rm core}_e - e V^{\rm core} = e^{\nu_{\rm 
crust}/2}\mu^{\rm crust}_e,
\end{equation}
where $\mu^{\rm core}_e = \mu_e(R_{\rm core})$, $e V^{\rm core}=e V(R_{\rm 
core})$, and $\mu^{\rm crust}_e = \mu_e(R_{\rm core} + \Delta r)$, and 
$e^{\nu_{\rm crust}} \simeq e^{\nu_{\rm core}}$, with the effective potential 
for the electrons $\mathcal{V}_e=-e V$, namely the electron Coulomb potential 
energy (since electrons are blind to the strong force). 

We have used in Eq.~(\ref{eq:Efermicrust}) the fact that no Coulomb potential 
appears in the crust and our previous results which show that the transition 
occurs at distance-scales $\Delta r \sim \hbar/(m_e c) \ll R_{\rm core}$, 
leading to a negligible gradient of the gravitational potential in such a 
region. This equation tells us that the Coulomb potential gap created between 
the charged core and the crust is of the order of $\Delta eV \sim \Delta 
\mu_e=\mu^{\rm core}_e-\mu^{\rm crust}_e$, which for $\mu^{\rm crust}_e\ll 
\mu^{\rm core}_e$ becomes of the order of $m_\pi c^2$ \cite{rotondo11d}. Under 
such conditions, there is an electric field in the transition $E \sim \Delta 
V/\Delta r \sim (m_\pi/m_e) E_c$.

In the transition region both the electron chemical potential and the density 
decrease (see Fig.~\ref{fig:Model}), namely $\mu^{\rm crust}_e < \mu^{\rm 
core}_e$ and $\rho_{\rm crust}<\rho_{\rm core}$. The electron chemical potential at the base of the crust $\mu^{\rm crust}_e$ is 
an increasing function of the crust base density $\rho_{\rm crust}$. Hence, for 
increasing values of $\rho_{\rm crust}$ the Coulomb potential gap is reduced 
leading to lower values of the electric field in the interface. The properties 
of this interface, such as its surface energy, the electric field, and the 
consequent Coulomb energy, were recently explored in \cite{2014PhRvC..89c5804R} 
up to the limit of locally neutral configurations for which $\rho_{\rm crust}$ 
is close to the nuclear saturation value. It was there shown how (see, e.g., 
Fig.~4 in that reference), indeed, the Coulomb energy is a decreasing function 
of the crust matching density, $\rho_{\rm crust}$.

All the above implies that for each central density, a family of core-crust 
interfaces exists. Correspondingly, for a given central density, there is an 
entire family of crusts with different mass and thickness. We analyze below the 
thermal evolution of neutron stars for different values of the crust density 
$\rho_{\rm crust}$, all the way up to approaching the limit of locally neutral 
configurations, which occurs for $\mu^{\rm crust}_e \to \mu^{\rm core}_e$, 
namely when $\rho_{\rm crust}\to\rho_{\rm nuc}$, approximately.

It is worth mentioning before closing this section, for the sake of completeness, scenarios in which the sharp transition 
becomes smoother. First of all let us briefly discuss the case in which the crust is in a fluid-like state. It is clear, following the above discussion, that in 
such a case the Wigner-Seitz cell construction does not apply any longer and 
macroscopic gravito-polarization effects, owing to gravitational segregation of 
the ions, should appear. This would change the afore discussed core-crust 
boundary matching conditions with consequences on the interface properties, 
likely reducing the interfacial electric field. This would lead to a smoother 
transition and the star would possibly develop a Coulomb potential and an 
electric field from the center all the way up the surface.

In addition, we would like to recall the consequences of requesting that more than one charge 
is conserved (e.g.~baryon and electric) within the Gibbs construction of the thermodynamic phase-transition. 
This case leads to the appearance of ``mixed-phases'' in between of the pure homogeneous phases, where essentially 
the two phases coexist, one over a background formed by the other, and in which a non-vanishing electric charge 
must exist forming charged Wigner-Seitz cells. Therefore, also in this case only the global charge neutrality of the 
system can be 
imposed (see, Refs.~\cite{1992PhRvD..46.1274G,1995PhRvC..52.2250G,1997PhRvC..56.2858C,1999PhRvC..60b5803G,2000PhRvC..62b5804C,2001PhR...342..393G}, 
for further details). The equilibrium pressure of the phases varies with the density creating a spatially extended 
and smoother phase-transition region of non-negligible thickness with respect to the star radius and 
without any density jump. However, the homogeneous pure phases in these treatments are still subjected 
to the condition of local charge neutrality, and so they do not account for the possible presence of 
interior Coulomb fields caused by gravito-polarization, as previously discussed.

\section{Thermal evolution equations} \label{sec:3}

The general relativistic equations of energy balance and energy transport for 
the description of the thermal evolution, in the spherically symmetric case 
treated here, read \citep[see 
e.g.][]{1977ApJ...212..825T,2006NuPhA.777..497P,TSURUTA1965}
\begin{align}
 \frac{\partial{(L e^{\nu})}}{\partial{r}}&=-\frac{4\pi r^2}{\sqrt{1-2m/r}} 
\left[\epsilon_\nu e^{\nu} + c_v  \frac{\partial{(T e^{\nu /2})}}{\partial{t}} 
\right],
\label{eq:eq1}
\\
\frac{L e^{\nu}}{4 \pi r^2 \kappa}&=\sqrt{1-2m/r} \, \frac{\partial{(T 
e^{\nu/2})}}{\partial{r}}.
\label{eq:eq2}
\end{align}
Eqs.~(\ref{eq:eq1}--\ref{eq:eq2}) depend on the structure of the star through 
the variables $r$, $m(r)$, $\rho(r)$, and $\nu(r)$ that represent the radial 
distance, the mass function, the energy density, the general relativistic 
gravitational potential, respectively. The thermal variables are represented by 
the interior temperature $T(r,t)$, the luminosity $L(r,t)$, neutrino emissivity 
$\epsilon_\nu(r, T )$, thermal conductivity $\kappa(r,T)$ and specific heat per 
unit volume $c_v(r,T)$.

The boundary conditions of Eqs.~(\ref{eq:eq1}--\ref{eq:eq2}) are determined by 
the luminosity at the center and at the surface. The luminosity vanishes at the 
stellar center, i.e.~$L(r=0)=0$,  since at this point the heat flux is zero. At 
the surface, the luminosity is defined by the relationship between the mantle 
temperature, which we denote to as $T_b$, and the temperature outside of the 
star, $T(r=R)=T_s$ 
\citep{2004ApJS..155..623P,Potekhin1997a,Yakovlev2004,Page2009}.

As for the thermal processes taken into account, in the core we consider the 
direct\footnote{Allowed by energy-momentum conservation only if the particle 
fraction satisfies the triangle inequality $P^F_n< P^F_p+P^F_e$, where 
$P^F_{n,p,e}$ are the Fermi momenta of neutrons, protons, and electrons, 
respectively.} and modified Urca processes, and the Bremsstrahlung process among 
the nucleons. In the crust we have plasmon decay, $e^- e^+$ pair annihilation, 
electron-nucleus and electron-nuclei Bremsstrahlung. Heat capacity and thermal 
conductivity follow their traditional formulation as described in 
\cite{Yakovlev2004} and references therein.


It is important to notice that pairing effects are extremely important for the 
cooling of a neutron stars  \citep[see e.g.][for a recent study of the effects 
of pairing in the thermal evolution of compact stars]{Page2011,Yakovlev2011}. 
However, for the purposes of this work, that is to investigate the thermal 
properties of globally neutral neutron stars with relation to their possible 
different crust thickness, we focus our attention in the case in which pairing 
effects are ``turned off'', as to obtain a better comprehension of the effects 
of the crust thickness to the cooling. For the sake of completeness we will 
perform a preliminary study in which we consider a conservative scheme for 
neutron and proton pairing. Clearly, once we obtain a better comprehension of the thermal evolution of 
globally neutral neutron stars we intend to augment this study by including  
more sophisticated processes, such as more sophisticated pairing and rotation 
effects \citep{2012PhRvD..85j4019N,Negreiros2011}

\section{Cooling curves and thermal relaxation time}\label{sec:4}

\begin{figure}[!hbtp] 
 \centering\includegraphics[width=\hsize,clip]{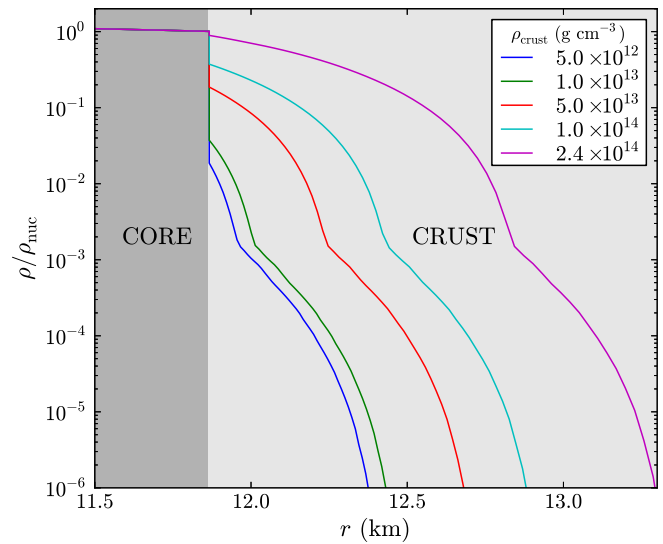}
 \caption{Density profiles of globally neutral neutron star with mass $M\approx 
1.4~M_\odot$ for selected values of the density at the base of the crust, 
$\rho_{\rm crust}$. Notice that at the transition, the density of the core is 
that of the nuclear saturation density, $\rho_{\rm nuc}$.}
 \label{fig:density}
\end{figure}

We turn now to the general properties of the cooling curves of globally neutral 
neutron stars. 
Exploring the freedom allowed by the global charge neutrality approach (namely 
the value of $\rho_{\rm crust}$), we investigate the thermal evolution of stars 
with different structures. Each neutron star study has the same gravitational 
mass ($M\approx 1.4~M_\odot$) with different values for $\rho_{\rm crust}$, 
which in turn is reflected in the thickness of the crust. In other words the 
core of these objects are largely the same, their crust, however, are 
increasingly thinner for smaller values of the density at its base, $\rho_{\rm 
crust}$. This is illustrated in Fig.~\ref{fig:density}. We note that the 
configuration whose $\rho_{\rm crust}=2.4\times 10^{14}$~g~cm$^{-3}$ has 
approximately the structure of a locally neutral neutron star, so this 
configuration should exhibit traditional cooling properties known in the 
literature.

We computed the cooling curves by integrating numerically the energy balance and 
transport equations (\ref{eq:eq1}--\ref{eq:eq2}) for the configurations shown in 
Fig.~\ref{fig:density}. 

\begin{figure}[!hbtp]
  \centering\includegraphics[width=\hsize,clip]{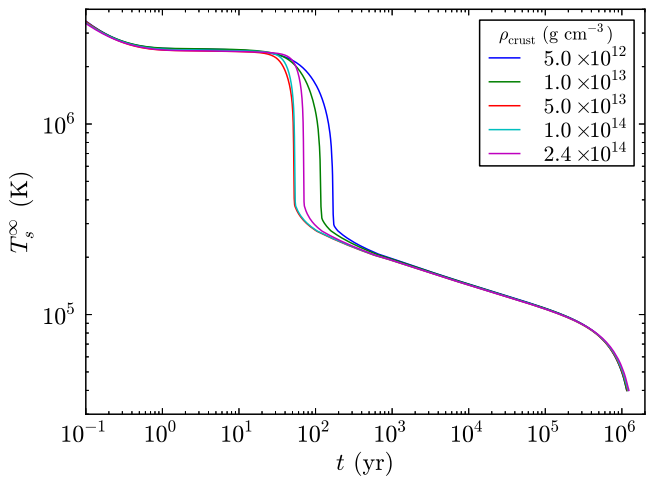}
 \caption{Surface temperature at infinity $T_s^\infty$ as a function of time $t$ 
in yr for the neutron star configurations shown in Fig.~\ref{fig:density}.}
 \label{fig:difmass}
\end{figure}

In Fig.~\ref{fig:difmass} we show the surface temperature as observed at 
infinity, as a function of time $t$ in yr for the neutron star configurations 
shown in Fig.~\ref{fig:density}. 

Fig.~\ref{fig:difmass} shows us that the different configurations have 
qualitatively the same behavior. However, the thermal relaxation time, $t_w$ 
\citep{2001MNRAS.324..725G}, of each is different. This quantity is defined as 
the time at which the star becomes isothermal. Such time is signaled by  a 
temperature drop in the surface temperature. The intensity of such drop will 
depend on whether or not fast neutrino processes (such as the DU) are taking 
place in the core. For stars younger than $t=t_w$, the crust of the star is 
hotter than the core and so heat flows from the crust to the core, or 
equivalently as a cooling front propagating from the core to the crust. The time 
$t_w$ is therefore the time it takes for  cooling front to reach the star 
surface.

\begin{figure}[!hbtp] 
 \centering\includegraphics[width=\hsize,clip]{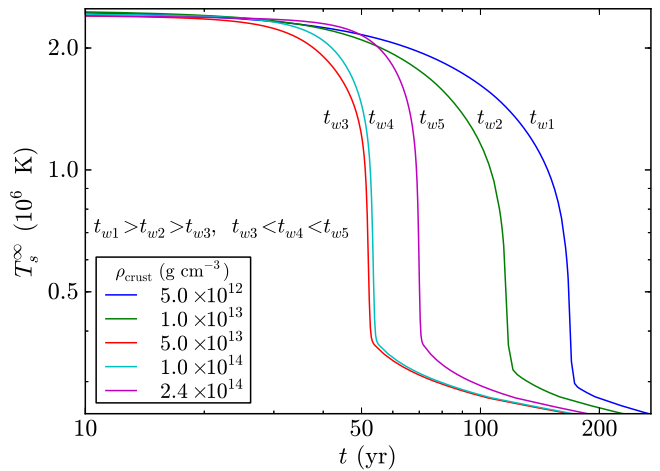}
  \caption{Enlargement of the evolution of the surface temperature around its 
drop at the end of the thermal relaxation phase, for the neutron stars shown in 
Fig.~\ref{fig:density}.}
  \label{fig:relaxtime}
\end{figure}

For traditional neutron stars, namely under the assumption of local charge 
neutrality, \cite{1994ApJ...425..802L} it is found by numerical simulations that 
$t_w$ is longer for thicker crusts, following approximately $t_w\propto \Delta 
R_{\rm crust}^{1.4-1.8}$, where $\Delta R_{\rm crust}$ is the crust thickness. 
The Fig.~\ref{fig:relaxtime} is an enlargement of Fig.~\ref{fig:difmass} around 
the temperature drop at the end of the thermal relaxation phase. As illustrated 
in Fig.~\ref{fig:difmass}, we see that for the neutron stars we are studying, 
the aforementioned behavior is only valid for stars whose $\rho_{\rm crust} \geq 
5\times 10^{13}$~g~cm$^{-3}$. For stars with $\rho_{\rm crust}\leq 5\times 
10^{13}$~g~cm$^{-3}$, however, we observe the opposite behavior, that is, an 
increase in $t_w$ for thinner crusts. In order to understand such behavior we 
need to realize two things: 1: The thickness of the crust contributes to 
increase $t_w$, since for a thicker crust the cooling front needs to ``travel'' 
a 
larger distance; and 2: For crusts with larger $\rho_{\rm crust}$ we have more 
intense emission of neutrinos from the crustal region. In the case of 
traditional neutron stars the range of densities covered by the crust is roughly 
the same, thus if the crust thickness is reduced one naturally obtains a smaller 
$t_w$ (since the crustal neutrino emission stays the same while the crust is 
thinner). In our study however, we consider different crust thickness 
\textit{and $\rho_{\rm crust}$}, thus while initially (when the variation of  
$\rho_{\rm crust}$ is small) the crustal neutrino emission is roughly the same 
(while the thickness is reduced), we have the reduction of $t_w$ (as in 
traditional neutron stars). However, when $\rho_{\rm crust}$ is significantly 
reduced, the neutrino emitting region is also significantly reduced. Therefore, 
we have two competing processes: the increase of $t_w$ due to the reduction of 
neutrino emission and the decrease of $t_w$ due to the geometrical reduction of 
the crust thickness. We 
have found that for  $\rho_{\rm crust}\lesssim 5\times 10^{13}$~g~cm$^{-3}$ the 
former wins and we have an overall increase of $t_w$ for such stars.  

\section{Pairing effects}\label{sec:5}

For completeness we have also considered pairing in the interior of neutron 
star. As mentioned before, neutron and proton pairing is fundamental for the 
temperature evolution of the object. In this study, with the intention of not 
deviating from the main goal (i.e. the investigation of the core-crust 
transition in the thermal evolution) we chose a conservative pairing scheme, in 
which we have neutron singlet ($^1S_0$) pairing in the crust (for the regions 
above neutron drip density), neutron triplet ($^3P_2$) pairing and proton 
singlet ($^1S_0$) pairing for the core region. We notice that in our 
conservative approach we do not have pairing extending to high densities regions 
of the core. 
\begin{figure}[!hbtp]
  \centering\includegraphics[width=\hsize,clip]{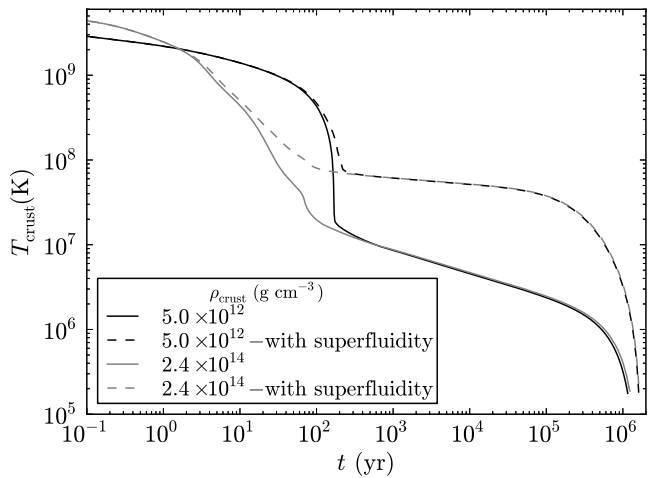}
 \caption{Same as in Fig.~\ref{fig:difmass} but for neutron stars with proton 
and neutron pairing.}
 \label{fig:cool_SF}
\end{figure}

We summarize our results in Fig.~\ref{fig:cool_SF}, where the thermal evolution 
of neutron stars with proton and neutron pairing is compared to the results 
previously discussed. For this analysis we have only used the two extreme 
core-crust transitions, with all the other results lying in between. As 
illustrated in Fig.~\ref{fig:cool_SF} the inclusion of pairing has its 
traditional effect of slowing the thermal evolution of the star, although the 
objects with different core-crust configurations exhibit, qualitatively, the 
same behavior as discussed before.


\section{Discussion}\label{sec:6}

In this work we have investigated the thermal properties of neutron stars 
satisfying global, rather than local, charge neutrality conditions. Within this 
model one no longer has a contiguous transition from the core to the crust of 
the neutron star. Instead there is gap ($\Delta R \sim 10^{2 -3}$fm) filled by a 
distribution of electrons and an ultra-high electric field. Under this 
description the density at the base of the crust might be as low as the neutron 
drip density, which means that one may obtain neutron stars with very different 
crusts: thin for the models whose density at the crust base is low; or thick in 
the case of larger values of the crust base density; all the way to the limit of 
local charge neutrality (where there is a continuous transition from core to 
crust). Evidently the question of which description (thicker or thinner crust) 
arises. Since the crust plays a small role in the the mass of the neutron star, 
and the fact that observation of the radii are still far from coming to reality, 
we resort to the thermal evolution of neutron stars to obtain hints on how to 
constrain the crust thickness.

We have shown that the structure of the neutron stars satisfying global charge 
neutrality may potentially lead to specific signatures in the thermal evolution 
of the neutron star. In particular, it has been shown that the traditional 
proportionality between the relaxation time and the crust thickness is violated 
for stars whose densities at the crust base is lower than $\approx 5\times 
10^{13}$~g~cm$^{-3}$, in which case the opposite behavior occurs, namely  one 
obtains a longer relaxation time for thinner crust.  The reason for this is the 
reduction of the crustal neutrino emission (due to the overall reduction of the 
crust density), which  contributes to keeping the crust warm, compensating the 
speed up even though the reduction of the crust size tend to speed up cooling. 
It is important to notice that for densities higher than this value the 
traditional behavior is restored, in which the relaxation time increases with 
the crust thickness.
These subtle effects are potentially observable in the thermal relaxation phase 
of the neutron star, which however has been observationally evasive until now. 
The observations of this early evolution of the neutron star might therefore 
probe the properties of the core-crust transition and thus may aid us in testing 
the underlying theory of global charge neutrality. We also call the attention to 
the fact that in the case in which the density at the base of the crust is equal 
to the neutron drip density ($\rho_{\rm drip}\approx 4.3\times 
10^{11}$~g~cm$^{-3}$), we have a situation that is similar to that of quark 
stars. Such hypothetical objects are composed of absolutely stable quark matter, 
and thus, due to charge neutrality at the surface of the quark core an 
ultra-high electric field arises filling the region between the core and a thin 
crust made up of ordinary matter, much like in the model describe in this work. 
Due to the similarities between these two models it is interesting to 
investigate and compare the cooling behavior of compact objects described by 
each model. Such studies are currently under way.

\begin{acknowledgments}
We would like to thank the anonymous referee for the important suggestions which 
helped us to improve the presentation of our results. S.M.C. acknowledges the 
support given by the International Relativistic Astrophysics PhD Program through 
the Erasmus Mundus Grant 2010--1816 from EACEA of the European Commission, 
during which part of this work was developed. S.M.C. and J.A.R. also 
acknowledge the support by the International Cooperation Program CAPES-ICRANet 
financed by CAPES -- Brazilian Federal Agency for Support and Evaluation of 
Graduate Education within the Ministry of Education of Brazil. R. N. 
acknowledges financial support by FAPERJ and CNPq.
\end{acknowledgments}

 References                         

\bibliographystyle{apsrev}
\bibliography{biblio}


\end{document}